
\magnification=1200
\baselineskip=18pt

\centerline{\bf SOME NEW SOLUTIONS OF YANG-BAXTER EQUATION}

\bigskip

\centerline{by}

\bigskip

\centerline{Susumu Okubo}
\centerline{Department of Physics and Astronomy}
\centerline{University of Rochester}
\centerline{Rochester, NY 14627}

\bigskip
\bigskip
\bigskip
\bigskip
\bigskip
\bigskip
\bigskip
\bigskip

\noindent {\bf Abstract}

\medskip

We have found some new solutions of both rational and trigonometric types
by rewriting  Yang-Baxter equation as a triple product equation in a
vector space of matrices.

\vfill\eject

The Yang-Baxter equation (YBE)
$$\sum^N_{j,k,\ell =1} R^{jk}_{a_1b_1} (\theta) R^{\ell a_2}_{kc_1}
(\theta^\prime) R^{c_2b_2}_{j \ell} (\theta^{\prime \prime}) =
\sum^N_{j,k,\ell =1} R^{\ell j}_{b_1c_1} (\theta^{\prime \prime} )
R^{c_2 k}_{a_1 \ell} (\theta^\prime) R^{b_2a_2}_{kj} (\theta)\eqno(1)$$
$$\theta + \theta^{\prime \prime} = \theta^\prime \eqno(2)$$
appears in many subjects ranging from statistical physics [1], exactly
solvable 2-dimensional field theories [1], and braid-group ([2]
and [3]), as well as the quantum group ([1] and [4]).  Let $V$ be a
$N$-dimensional vector space with a symmetric bilinear non-degenerate form
$<x|y>\ =\ <y|x>$.  For a fixed basis $e_1,\ e_2,\ \dots,\ e_N$
 of $V$, we set
$$g_{jk} = g_{kj} = \ < e_j |e_k> \eqno(3)$$
with its inverse $g^{jk}$.  We raise and lower indices as usual in terms of
these metric tensors as
$$e^j = \sum^N_{k=1} g^{jk} e_k \quad . \eqno(4)$$
We introduce [5] two $\theta$-dependent triple products by
$$\eqalignno{\big[ e^c , e_a , e_b \big]_\theta &= \sum^N_{d=1} e_d R^{dc}_{ab}
(\theta) \quad , &(5a)\cr
\noalign{\vskip 4pt}%
\big[ e^d , e_b , e_a \big]^*_\theta &= \sum^N_{c=1}  R^{dc}_{ab}
(\theta) e_c  &(5b)\cr}$$
so that we have
$$R^{dc}_{ab} (\theta) =\ <e^d | \left[ e^c , e_a , e_b \right]_\theta
>\ = \ <e^c | \left[ e^d , e_b , e_a
\right]^*_\theta > \quad . \eqno(6)$$
Then as we noted in [5], YBE can be rewritten as a triple product equation
$$\eqalign{\sum^N_{j=1} \big[v, [u, e_j &, z ]_{\theta^\prime} , [e^j ,
x , y]_\theta \big]^*_{\theta^{\prime \prime}}\cr
&= \sum^N_{j=1} \big[u, [v, e_j , x ]^*_{\theta^\prime} , [e^j ,
z , y]^*_{\theta^{\prime \prime}} \big]_\theta \quad . \cr}\eqno(7)$$
As a matter of fact, if we identify $x= e_{a_1},\ y = e_{b_1}$,
$z = e_{c_1}$, $u = e^{a_2}$, and $v = e^{c_2}$ in Eq. (7), and if we note
Eqs. (5), then we can readily verify that Eq. (7) will reproduce Eq. (1).
Similarly, we have the validity of
$$<u| [v,x,y]_\theta >\ =\ <v |[u,y,x]^*_\theta > \eqno(8)$$
for any $u,\ v,\ x,\ y\ \epsilon \ V$ in view of Eq. (6).

In reference [5], some solutions of Eq. (7) for the case of
$[x,y,z]^*_\theta = [x,y,z]_\theta$ have been found for some triple systems
including the case of the octonionic solution of de Vega and Nicolai [6].
The purpose of this note is to present another simpler solutions in terms
of $n \times n$ matrices.  Let $V$  now be a vector space consisting of all
$n \times n$ matrices with $N = n^2$, i.e.,
$$V = \{ x | x = n \times n\ {\rm matrix}\} \eqno(9)$$
and set
$$<x|y>\ = \ {\rm Tr}\ (xy) \quad . \eqno(10)$$
The completeness condition of the space $V$ can then be expressed as
$$\sum^N_{j=1} e_j x e^j = ({\rm Tr}\ x)\ {\bf 1} \eqno(11)$$
for any $x\ \epsilon\ V$, where {\bf 1} stands for the unit $n \times n$
matrix.

Following the reference [5], we seek a solution with the ansatz of
$$\eqalignno{[x,y,z]_\theta &= P_1 (\theta)
yzx + P_2 (\theta) xzy + A(\theta) <y|z> x + C(\theta)
<z|x>y \quad , &(12a)\cr
[x,y,z]_\theta^* &= P_2 (\theta)
yzx + P_1 (\theta) xzy + A(\theta) <y|z> x + C(\theta)
<z|x>y \quad , &(12b)\cr}$$
which satisfy the constraint Eq. (8).  Here, $P_1(\theta),\ P_2(\theta),\
 A(\theta),\ {\rm and}\ C(\theta)$ are some functions of $\theta$ to be
determined.  Also, the products $yzx$ and $xzy$ in Eqs. (12) represent the
standard associative matric products in $V$.  We insert the expression Eqs.
(12) into both sides of Eq. (7) and note the validity of Eq. (11).  This
yields the following equation:
$$\eqalign{O &= \sum^N_{j=1} \big\{ \big[ v, [u, e_j , z]_{\theta^\prime} ,
[e^j , x, y]_\theta \big]^*_{\theta^{\prime \prime}}\cr
&\qquad - [u, [v,e_j ,x]^*_{\theta^\prime} , [e^j ,z
,y ]^*_{\theta^{\prime \prime}} \big]_\theta \big\}\cr
&= K_0 (uxyzv - vzyxu)\cr
&\qquad + K_1 (yzuxv - vxuzy) - \hat K_1 (yxvzu - uzvxy)\cr
&\qquad + K_2 \{ <z|u> yxv\  -\ <v|x>uzy\} - \hat K_2 \{ <x|v>yzu\  -\
<u|z>vxy\}\cr
&\qquad + K_3 \{ <x|y>uzv\  -\ <y|xvz>u\} - \hat K_3
\{ <z|y>vxu\  - \ <y|zux>v\}\cr
&\qquad + K_4 \{ <y|x>vzu\  - \ <y|zvx>u\} - \hat K_4 \{
<y|z>uxv \  -\ <y|xuz>v\}\cr
&\qquad + K_5 <x|y><z|u>v - \hat K_5 <z|y><x|v>u \quad ,\cr} \eqno(13)$$
where we have set for simplicity
$$\eqalign{K_0 &= P^{\prime \prime}_2 A^\prime P_1 - P^{\prime \prime}_1
A^\prime P_2 \quad ,\cr
K_1 &= P^{\prime \prime}_2 P^\prime_1 C - P^{\prime \prime}_1
P^\prime_2 C\cr
\hat K_1 &= C^{\prime \prime} P^\prime_2 P_1 - C^{\prime \prime}P^\prime_1
P_2\cr
K_2 &= P^{\prime \prime}_2 P^\prime_1 P_2 + P^{\prime \prime}_2 C^\prime
C + C^{\prime \prime} C^\prime P_2 - C^{\prime \prime} P_2 C +
n P^{\prime \prime}_2 C^\prime P_2 \quad ,\cr
\hat K_2 &= P^{\prime \prime}_1 P^\prime_2 P_1 + C^{\prime \prime} C^\prime
P_1 + P^{\prime \prime}_1 C^\prime C - C^{\prime \prime} P^\prime_1 C +
n P^{\prime \prime}_1 C^\prime P_1 \quad ,\cr
K_3 &= P^{\prime \prime}_2 P^\prime_2 P_1 + P^{\prime \prime}_2 A^\prime
A + C^{\prime \prime} P^\prime_2 A - C^{\prime \prime} A^\prime
P_2  +
n P^{\prime \prime}_2 P^\prime_2 A \quad ,\cr
\hat K_3 &= P^{\prime \prime}_2 P^\prime_1 P_1 + A^{\prime \prime} A^\prime
P_1 + A^{\prime \prime} P^\prime_1 C - P^{\prime \prime}_1 A^\prime
C  +
n A^{\prime \prime} P^\prime_1 P_1 \quad ,\cr
K_4 &= P^{\prime \prime}_1 P^\prime_1 P_2 + P^{\prime \prime}_1 A^\prime
A + C^{\prime \prime} P^\prime_1 A - C^{\prime \prime} A^\prime
P_1  +
n P^{\prime \prime}_1 P^\prime_1 A \quad ,\cr
\hat K_4 &= P^{\prime \prime}_1 P^\prime_2 P_2 + A^{\prime \prime} A^\prime
P_2 + A^{\prime \prime} P^\prime_2 C - P^{\prime \prime}_2 A^\prime
C  +
n A^{\prime \prime} P^\prime_2 P_2 \quad ,\cr
K_5 &= P^{\prime \prime}_2 P^\prime_1 A + P^{\prime \prime}_1 P^\prime_2
A + A^{\prime \prime} P^\prime_1 P_2 + A^{\prime \prime} P^\prime_2
P_1  +
 P^{\prime \prime}_2 C^\prime P_1  + P^{\prime \prime}_1
C^\prime P_2 \cr
&\qquad + n \big\{ P^{\prime \prime}_2 C^\prime
A + P^{\prime \prime}_1 C^\prime A + A^{\prime \prime} P^\prime_1
A  +
A^{\prime \prime} P^\prime_2 A + A^{\prime \prime} C^\prime
P_1 + A^{\prime \prime} C^\prime P_2 \big\} \cr
&\qquad + C^{\prime \prime} C^\prime
A - C^{\prime \prime} A^\prime C + A^{\prime \prime} C^\prime
C +
A^{\prime \prime} A^\prime A + n^2 A^{\prime \prime} C^\prime
A \quad , \cr
\hat K_5 &= A^{\prime \prime} P^\prime_2 P_1 + A^{\prime \prime} P^\prime_1
P_2 + P^{\prime \prime}_1 P^\prime_2 A + P^{\prime \prime}_2 P^\prime_1
A  +
 P^{\prime \prime}_2 C^\prime P_1  + P^{\prime \prime}_1
C^\prime P_2 \cr
&\qquad + n \big\{ A^{\prime \prime} C^\prime
P_1 + A^{\prime \prime} C^\prime P_2 + A^{\prime \prime} P^\prime_2
A  +
A^{\prime \prime} P^\prime_1 A + P^{\prime \prime}_2 C^\prime
A + P^{\prime \prime}_1 C^\prime A \big\} \cr
&\qquad + A^{\prime \prime} C^\prime
C - C^{\prime \prime} A^\prime C + C^{\prime \prime} C^\prime
A +
A^{\prime \prime} A^\prime A + n^2 A^{\prime \prime} C^\prime
A \quad . \cr}\eqno(14)$$
Here, $P^{\prime \prime},\ P^\prime, \ {\rm and}\ P$ for example stand for
$$P = P(\theta)\quad , \quad P^\prime = P (\theta^\prime ) \quad , \quad
P^{\prime \prime} = P(\theta^{\prime \prime} ) \quad . \eqno(15)$$
We note that $\hat K_j \ (j = 1,2,3,4,5)$ is the same function as $K_j$
except for the interchanges of $\theta \leftrightarrow
\theta^{\prime \prime}$ and $P_1 \leftrightarrow P_2$.
The YBE can be satisfied, if we have
$$K_0 = K_1 = \hat K_1 = K_2 = \hat K_2 = K_3 = \hat K_3 = K_4 = \hat K_4
= K_5 = \hat K_5 = 0 \quad . \eqno(16)$$
We can solve these eleven coupled function equations as in [5] and [6] to
find the following trigonometric solutions, assuming that at least one of
$P_1 (\theta)$ and $P_2 (\theta)$ is not identically zero:

\medskip

\noindent \underbar{Solution (I)}

\medskip

We have $P_1 (\theta) = P_2 (\theta)$.  Setting
$$\lambda = {1 \over 2}\ \left( n \pm \sqrt{n^2 - 4} \right) \quad ,
\eqno(17)$$
the solution is given by
$$\eqalignno{{A (\theta ) \over P_1 (\theta)} &= - {\lambda^2 e^{k\theta} -
\beta \over \lambda (e^{k \theta} - \beta)} \quad , &(18a)\cr
\noalign{\vskip 4pt}%
{C (\theta ) \over P_1 (\theta)} &= - { e^{k\theta} - \lambda^2
 \over \lambda (e^{k \theta} - 1)} \quad , &(18b)\cr}$$
where $\beta$ can assume two possible values of $\lambda^2$ or
$- \lambda^4$, and $k$ is an arbitrary constant including the value of
$k = \pm \infty$.

\medskip

\noindent \underbar{Solution (II)}

\medskip

$$P_2 (\theta) = 0 \quad , \quad {C (\theta) \over P_1 (\theta)} = {n \over
e^{k \theta} - 1} \quad , \quad
{A (\theta) \over P_1 (\theta)} = {n e^{k \theta} \over
(n^2 -1) - e^{k \theta}} \quad . \eqno(19)$$

\medskip

\noindent \underbar{Solution (III)}

\medskip

$$P_1 (\theta) = 0 \quad , \quad {C (\theta) \over P_2 (\theta)} = {n \over
e^{k \theta} - 1} \quad , \quad
{A (\theta) \over P_2 (\theta)} = {n e^{k \theta} \over
(n^2 -1) - e^{k \theta}} \quad . \eqno(20)$$
In Eqs. (19) and (20), $k$ is again an arbitrary constant including the
case of $k = \pm \infty$.

We remark that these solutions satisfy the so-called crossing realtion
([2] and [7]) which can be expressed as
$${1 \over P_1 (\overline \theta)}\ [y,x,z]_{\overline \theta} =
{1 \over P_1 (\theta)} \ [x,y,z]_\theta \eqno(21)$$
for example for solutions (I) and (II), where $\overline \theta$ is related
to $\theta$ by
$$k (\overline \theta + \theta) = \cases{\log \beta &for (I)\cr
\log (n^2 -1) &for (II) \quad .\cr} \eqno(22)$$
They also satisfy the unitarity relation.  To see it, we introduce
$\underline{R} (\theta)$ and $\underline{R}^* (\theta)\ :\
V \otimes V \rightarrow V \otimes V$ by
$$\eqalignno{\underline{R}
 (\theta) e_a \otimes e_b &= \sum^N_{j,k=1} R^{kj}_{ab}
 (\theta) e_j \otimes e_k &(23a)\cr
\underline{R}^* (\theta) e_a \otimes e_b &= \sum^N_{j,k=1} R^{jk}_{ba}
 (\theta) e_j \otimes e_k \quad . &(23b)\cr}$$
The relationship between $\underline{R} (\theta), \
\underline{R}^* (\theta)$, and triple products
 is given then by
 $$\eqalignno{\underline{R} (\theta) x \otimes y &= \sum^N_{j=1} e_j \otimes
\big[ e^j , x, y\big]_\theta =
\sum^N_{j=1} \big[ e^j, y, x \big]^*_\theta \otimes e_j \quad ,
&(24a)\cr
\underline{R}^* (\theta) x \otimes y &= \sum^N_{j=1} e_j \otimes
\big[ e^j , x, y\big]^*_\theta =
\sum^N_{j=1} \big[ e^j, y, x \big]_\theta \otimes e_j \quad .
&(24b)\cr}$$
Now, the unitarity relation for all solutions (I)-(III) is expressed in the
form of
$$\underline{R}^* (- \theta) \underline{R}(\theta) = \underline{R}
(\theta) \underline{R}^* (- \theta) = C(\theta) C(-\theta ) I_d
\eqno(25)$$
where $I_d$ is the identity map in $V \otimes V$.  Note expecially that we
have $\underline{R}^* (\theta) = \underline{R} (\theta)$ and
$[x,y,z]^*_\theta = [x,y,z]_\theta$ for the solution (I).

We have also found another solution of YBE when we replace Eqs. (12) now
by
$$\eqalign{[x,y,z]_\theta &= [x,y,z]^*_\theta\cr
&= P_1 (\theta) zxy + P_2 (\theta) yxz + B(\theta) <x|y>z + C(\theta) <z|x>
y \cr} \eqno(26)$$
which is consistent with Eq. (8).  Repeating the same procedure as before,
the solutions are now found to be of rational type given by

\medskip

\noindent \underbar{Solution (I)}

\medskip

$$P_2 (\theta) = \alpha P_1 (\theta) \quad , \quad {B(\theta) \over P_1 (
\theta)} = k \theta \quad , \quad {C(\theta) \over P_1 (\theta)} =
{\alpha \over k \theta}
 \eqno(27a)$$

\medskip

\noindent \underbar{Solution (II)}

\medskip

$$P_2 (\theta) = C(\theta) = 0 \quad , \quad {B(\theta) \over P_1 (\theta)}
= \beta + k\theta \eqno(27b)$$

\medskip

\noindent \underbar{Solution (III)}

\medskip

$$P_1 (\theta) = C(\theta) = 0 \quad , \quad {B(\theta) \over P_2
 (\theta)} = \beta + k\theta \eqno(27c)$$
for arbitrary constants $\alpha (\not= 0),\ \beta,\ {\rm and}\ k$.
However, we will not go into details of the calculations.

\medskip

\noindent {\bf \underbar{Acknowledgements}}

\medskip

\line{This \hfill paper \hfill is\hfill supported \hfill in \hfill
 part \hfill by \hfill the \hfill U.S. \hfill Department \hfill
 of \hfill  Energy \hfill Grant}

\noindent DE-FG-02-91ER40685.

\vfill\eject

\noindent {\bf \underbar{References}}

\medskip

\item{1.} M. Jimbo, Yang-Baxter Equations in Integrable Systems, (World
 Scientific, Singapore 1989).

\item{2.} C. N. Yang and M. L. Ge, Braid Group, Knot Theory, and Statistical
Mechanics, (World Scientific, Singapore 1989).

\item{3.} L. H. Kauffman, Knots and Physics (World Scientific, Singapore,
1991).

\item{4.} Y. I. Manin, Quantum Groups and Non-Commutative Geometry,
(University of Montreal Press, Montreal, 1988).

\item{5.} S. Okubo, Jour. Math. Phys. {\bf 34}, (1993) 3273, 3292.

\item{6.} H. J. de Vega and H. Nicolai, Phys. Lett. {\bf B244}, (1990)
 295.

\item{7.}  A. B. Zamolodchikov and Al. B. Zamolodchikov, Ann. Phys.
{\bf 120}, (1979) 253.

\end